\documentclass[10pt,letterpaper]{article}

\usepackage{amsmath,amssymb}

\usepackage{changepage}

\usepackage[utf8x]{inputenc}

\usepackage{textcomp,marvosym}

\usepackage{cite}

\usepackage{nameref,hyperref}

\usepackage{microtype}
\DisableLigatures[f]{encoding = *, family = * }

\usepackage[table]{xcolor}

\usepackage{array}

\usepackage{comment}
\usepackage{longtable}
\usepackage{subfig}
\usepackage{graphicx}
\usepackage{adjustbox}
\usepackage{verbatim}
\usepackage{color} 
\usepackage{multirow}
\usepackage{rotating}

\begin{document}


{\Large
\textbf\newline{Prior Knowledge based mutation prioritization towards causal variant finding in rare disease} 
}
\newline
\\
Vasundhara Dehiya\textsuperscript{1,2},
Jaya Thomas\textsuperscript{1,2},
Lee Sael\textsuperscript{1,2*}
\\
\bigskip
\textbf{1} Department of Computer Science, State University of New York Korea, Incheon, Republic of Korea
\\
\textbf{2} Department of Computer Science, Stony Brook University, Stony Brook, NY, USA
\\
\bigskip

%
%





* sael@cs.stonybrook.edu (LS)


\section*{Abstract}
How do we determine the mutational effects in exome sequencing data with little or no statistical evidence?
Can protein structural information fill in the gap of not having enough statistical evidence?
In this work, we answer the two questions with the goal towards determining pathogenic effects of rare variants in rare disease.
We take the approach of determining the importance of point mutation loci focusing on protein structure features.
The proposed structure-based features contain information about geometric, physicochemical, and functional information of mutation loci and those of structural neighbors of the loci. The performance of the structure-based features trained on 80\% of HumDiv and tested on 20\% of HumDiv and on ClinVar datasets showed high levels of discernibility in the mutation's pathogenic or benign effects: F score of 0.71 and 0.68 respectively using multi-layer perceptron.
Combining structure- and sequence-based feature further improve the accuracy: F score of 0.86 (HumDiv) and 0.75 (ClinVar).
Also, careful examination of the rare variants in rare diseases cases showed that structure-based features are important in discerning importance of variant loci.

\section*{Author summary}
We explore how well features that do not rely on sample sizes can determine the pathogenic effects of variant with the goals to correctly determine the pathogenic effects of rare variants in rare disease.
We examined the deterministic factor of loci-based features in classifying pathogenic variant verses benign variant focusing on the structure-based features as a way to be independent of the number of samples available for a disease.
Unlike the previous studies that tend to ignore structure-based features, we emphasize the structure-based features and enhance it by using information from multiple homolog structures and from structural neighbors.
As results, we can get up to F score of 0.86 using only the loci-based information and F score of 0.71 using only the loci-based structure-based features.
High F scores and empirical examination of rare variant in rare diseases mapped to structures shows that where the variant have occurred in structure is an important factor in determining pathogenic effects.

\section*{Introduction}
Rare diseases have been known to affect over 350 million people worldwide having over 7,000 different rare diseases 80\% of which are due to genetic factors~\cite{GlobalGenes}.
Compared to common and complex disease, rare disease are often heterogeneous, rare (low minor allele frequency less than 1\%), and are often hereditary~\cite{de2013background, chial2008mendelian}
Many rare diseases can be cared for if diagnosed early and optimally managed.
However, accurate diagnosis, adequate information, and ongoing targeted care are significant challenges in rare disease treatment.
Of many challenges, a quick and non-statistical method for identifying causal variant of rare diseases through mutation prioritization forms the basis for fast and accurate diagnosis as well as for providing effective treatment suggestions~\cite{ng2010exome, Boycott2013}.

The goal of variant (mutation) prioritization is to determine the relative importance of the variation in the context of disease mechanism.
The informatics approach for mutation prioritization considers two factors: the importance of the variant loci and the severity of the mutation itself.
The importance of the loci is a prior knowledge that is independent of the variation given the loci.
This means the importance of the loci, i.e. loci weights, can be trained and learned in the preprocessing step independent of disease types thus combined data from several diseases can be used.
On the other hand, the severity of the mutation rely on the information of the mutation compared to wild-type and often requires statistical measures for specific diseases of interest.
Although both factors are important in determining the effect of a variant, it will be difficult to determine to causality if the severity of variant is measure based on statistical means of specific disease for rare variant cases.
Thus it is meaningful to study the how well prior information alone can determine the pathogenic effect.
Thus, we propose to approach this challenge by separating out the features of determining the importance of loci and for determining the severity of mutation itself and focusing on determination of the importance of loci in terms of function.

Considering the source of information that characterizes loci in a genome sequence, the features, generated from the two factors previously mentioned, can be separated to sequence-based features and structure-based features.
Most sequence-based methods generate variant characteristics from protein and DNA sequences via multiple sequence alignment to determine the effect variants~\cite{ng2003sift, mathe2006computational}. Sequence-based features often include sequence conservation related scores such as substitutions scores for protein coding regions.
Divers type of features can be extracted from protein structure and they can be classified largely to physio-chemical properties, structure stability, and binding information.
Physiochemical properties include hydrophobicity, amino acid (AA) weight, AA volume, and isoelectric point \cite{jiang2007sequence}.
Structure stability properties include B-factor, secondary structure information, and region in Ramachandran plot \cite{Carter2009, adzhubei2013predicting}.
Binding site information includes the interaction of mutation site with other biomolecules \cite{mueller2015ball}.
Sequence-based features, mostly due to their abundance and easiness of usage, have been widely used in finding causal genes.
The strength of sequence-based methods is that it can be extended non-coding regions of the whole genome \cite{yuan2006fastsnp, adzhubei2010method, kircher2014general}.
On the other hand, structure features are limited to protein coding regions, however, provides more direct information about the function.
In fact, exome sequencing, sequencing of the protein coding areas, has been found to be effective means of detecting pathogenic genes as many alleles associated with hereditary diseases, including many rare diseases, have been discovered to alter protein coding nucleotides \cite{bamshad2011exome}.

Features used for pathogenic single nucleotide variant (SNV) prediction methods are generally sequence-based only or sequence and structure combined.
Although there have been several extensive studies for sequence-based features due to their abundance and easiness of usage, there has not been a rigorously extensive examination of structure-based features in the causal mutation predictions.
Many structure focused methods are limited in application as specific to cancer~\cite{Carter2009}, are more of an annotator~\cite{Gress2016}.
Ones that are more general, considers the structure-based features as minor features~\cite{adzhubei2010method, adzhubei2013predicting, Ng2003}.
We believe this misunderstanding of structure features being minor is due to the relatively limited number and coverage of structural data compared to the sequence as well as relatively careless processing of structural data which has more divergent information and is not due to the importance of structural information.
Our study here is show that structure-based features are important in large study.
The limitations such as low coverage and limited function information can be address via various computational prediction methods, such as structure prediction in presence of mutation \cite{Pi2013}, reconstruction of protein-protein interactions \cite{Lo2015}, ligand binding site predictions \cite{Sael2012a}, and stability predictions \cite{Worth2011}.

There are several works that show mutations on functional and structurally stabilizing sites are strongly related to impaired protein function~\cite{bamshad2011exome,tennessen2012evolution} that results in diseases.
Disease causing P53 structures contains mutations at key sites that are important for maintaining structural integrity~\cite{Bullock2000, Joerger2006} and/or contain mutations in the DNA binding region locally changing the binding affinity and altering the function~\cite{freed2012mutant,Lwin2007}.
Point mutations in the protein stabilizing regions have also been observed in other diseases including Alzheimer's, Prion, and Parkinson's disease~\cite{Zhang2013a}.
Mutations at binding sites in proteins are functionally important as they lead to changes in binding affinity between the protein and other bio-molecules.
The affected binding targets may be ions, other proteins, ligands, DNA or RNA.
A study of somatic mutations in DNA methyltransferase gene DNMT3A in acute monocytic leukemia found mutations at DNA binding, cofactor binding, protein-protein interaction sites and Histone H3 peptide binding sites \cite{yan2011exome}.
These mutations are crucial as they prevent protein complex formation thereby leading to impaired gene function.
Also, rare mutations in DNA-binding site in POT1 protein have been identified as causal for cutaneous malignant melanoma as these mutations changed protein folding and binding \cite{shi2014rare}.
Another study identified somatic mutations in the pre-mRNA binding region in splicing factor SF3B1 gene which is known to cause chronic lymphocytic leukemia \cite{quesada2012exome}.
These studies highlight the importance of structure-based information in causal mutation analysis in disease.

Again, it is difficult to obtain similar disease samples, patient's data alone may be the only means of determining the causal genes. Considering that the number of rare variant in rare diseases cases are inherently limited, it is important to be able to extract information from non-statistical means.
Protein structure provides direct information about the functional importance of loci and does not rely on statistical means for prioritizing the importance of mutation loci.
These structure-based features need to be extensively tested, especially when several empirical evidence demonstrate causal mutations at critical structural positions.
Our features are significant in that structure features considers loci information only, utilizes multiple homolog structures for a given sequence, and considers the features of structural neighbors.

\subsubsection*{Contribution}
The contributions of the paper are listed as follows:
\begin{itemize}
 \item Development of loci specific, multi-structural, structural-neighbor considering features for casual variant prediction.
 \item Extensive study of the performance of structure-based features in determining pathogenic effects in point mutations.
 \item Comparative study of loci specific features using various machine learning methods and PolyPhen2 \cite{adzhubei2010method}.
 \item Demonstrate soundness of generated features for rare variant rare disease casual mutation predictions.
\end{itemize}

\section*{Materials and methods}

\subsection*{Data sets}
We generate two data sets, i.e. HumDiv data set, ClinVar dataset, ClinVar Rare Variant Rare Disease data set to compare and test the performance of proposed features.
Initially, SNV information, including loci, along with the known functional impact of that variant is extracted from HumDiv database \cite{adzhubei2010method} and ClinVar database \cite{landrum2016clinvar} separately.
HumDiv database contains a total of 13,103 SNVs, which consists of 5,564 pathogenic SNVs that are known to causing human Mendelian diseases, and 7,539 benign variants that are generated from differences between human proteins and their closely related mammalian homologs (seqID \textgreater=95\%) \cite{adzhubei2010method}.
Among these, 11,422 have unique variant loci.
From the initial dataset, after filtering out SNV without proper protein structures mapped, 2,473 pathogenic and 2,992 benign (total of 5,465) SNVs remain and are used as HumDiv dataset.
The HumDiv dataset is used mainly for training the model and for comparing with a existing SNV classification method, PolyPhen2 \cite{adzhubei2010method}, which is a widely used method that includes structural information.

The ClinVar database contains an archive of human variations and phenotypes, with supporting evidence and is more up-to-date and carefully curated~\cite{landrum2016clinvar}.
From the initial database of 93,946 SNV entries retrieved Jan. 2016, we select 9,286 SNVs with review status 2 through 4 (highest);
the review status $2$ is `SNP submitted with criteria provided, two or more submitters with no conflicts'; status $3$ is `reviewed by an expert panel'; status $4$ is `practice guideline'.
From the 9,286 SNVs, we selected 6,446 loci labeled as either likely benign or benign and likely pathogenic or pathogenic without conflicting labels.
As the final step, SNPs without protein structure mapping was removed resulting in a small set of 438 SNV loci, with 286 pathogenic and 152 benign labels in 101 genes.
The ClinVar dataset is used to test the model learned with HumDiv in the new dataset.

Rare variants in ClinVar are identified using Ensembl Variant Effect Predictor tool \cite{mclaren2016ensembl}.
This identified 402 rare variants in the ClinVar dataset.
An SNV is considered to be associated with a rare disease if one or more of the diseases label in ClinVar belong to the list of rare diseases, which is obtained through Global Genes\cite{GlobalGenes} that lists 6,537 rare diseases, as of August 6, 2017.
Among these rare variants, 117 benign rare variants and 163 pathogenic rare variants that are identified to be associated with a rare disease are used as ClinVar Rare Variant Rare Disease dataset.

\subsection*{Features to loci mapping}
\begin{figure}[h!]
 \centering
 \includegraphics[width = 10cm]{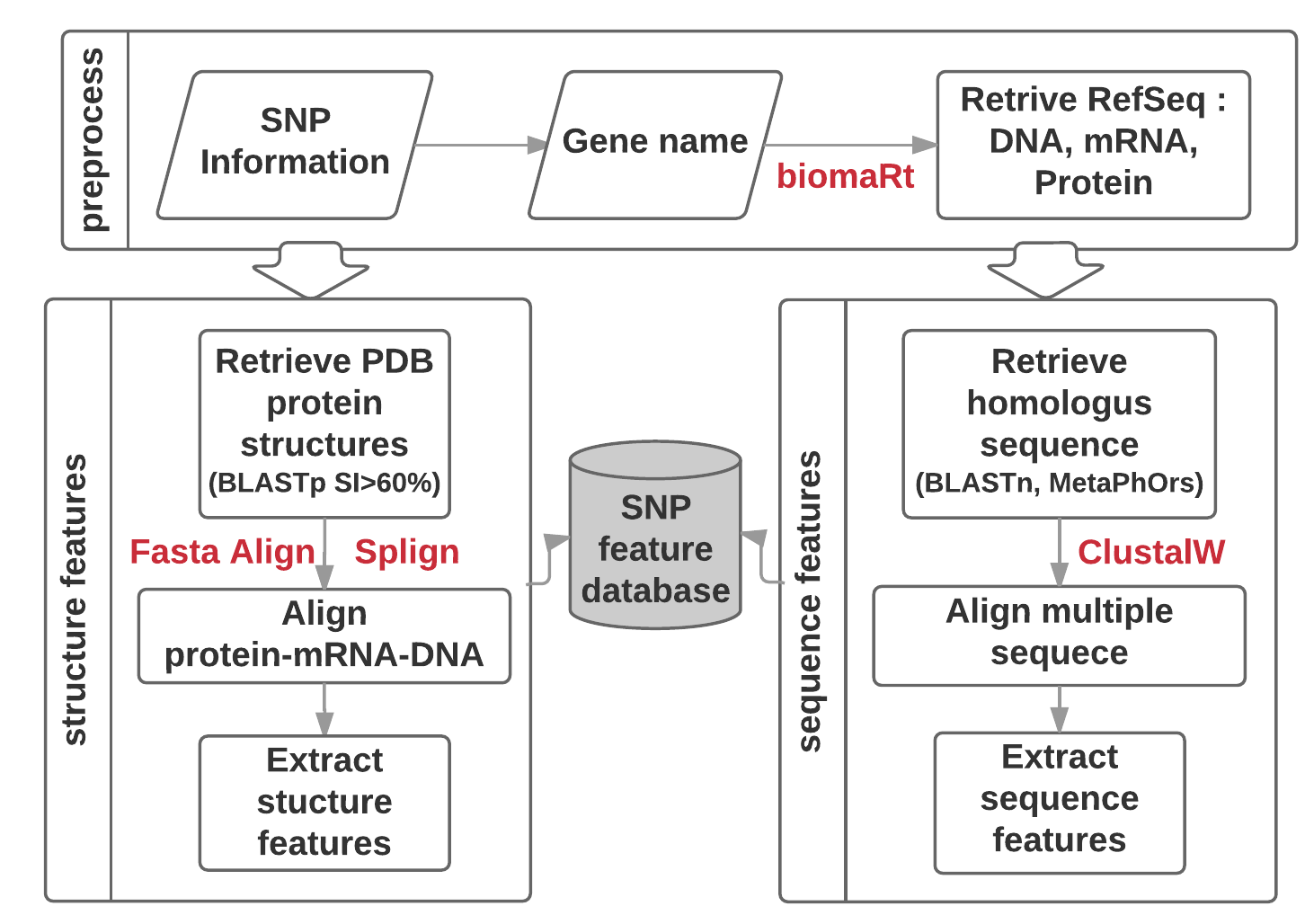}
 \caption{Loci specific feature extraction pipeline.}
 \label{fig:flowchart2}
\end{figure}

In order to extract loci specific features, the first step is to retrieve reference DNA, mRNA, and protein sequences of the corresponding loci.
To do this, we identify genes associated with the variant in the DNA loci represented as chromosome number and offset value and use biomaRt \cite{durinck2005biomart} library to retrieve reference sequences for DNA (DNA-RefSeq), mRNA (NM-RefSeq) and protein amino acid sequence (NP-RefSeq) from Ensemble database~\cite{aken2016ensembl} corresponding to genome reference HG38 for each dataset.
DNA, mRNA, and protein sequences are then aligned to map the features back the variant loci.
Fig \ref{fig:flowchart2} shows the steps involved in the feature generation process utilizing the retrieved sequences and steps is described in detail in the following feature extraction process description.

\subsection*{Extracting structure-based features}
Given a variant locus, the first step of structure-based feature extraction process is to retrieve protein structure for the locus.
To do this, we retrieve a list of protein structures (PDBIDs) of corresponding NP-RefSeq using protein BLAST \cite{altschul1990basic} with homology threshold of 60\% sequence identity.
If structures are found, the corresponding Protein Data Base (PDB) files are downloaded from RSCB \cite{berman2003announcing} database.
From the PDB structures, PDB sequences are directly extracted and pair-wise global aligned using a dynamic programming based aligner, i.e., FASTA align \cite{pearson1988improved}), to the protein reference sequence (NP-RefSeq).
Corresponding reference mRNA sequence (NM-RefSeq) of the NP-RefSeq is then aligned to reference DNA (NC-RefSeq) using Splign \cite{kapustin2008splign}, which allows alignment even in the presence of splice sites.
Finally, the structure-based features are generated and mapped to the variant locus.

\subsubsection*{Structural neighborhood}
\begin{figure}[h!]
 \centering
 \includegraphics[width=60mm]{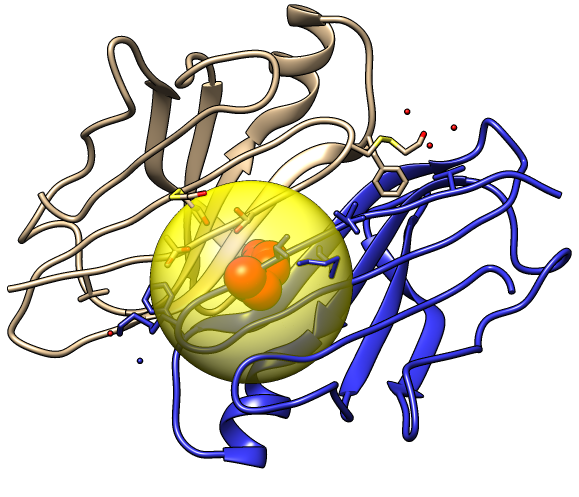}
 \caption{{\bf Structural neighbourhood} Yellow sphere depicts a radius of 9\AA ~and all amino acids residues found within this region are considered as neighbours of the query site. PDBID:1IJN centered at 116.A}
 \label{fig:neighbourhood}
\end{figure}

Features generated from structural neighboring are one of the novel contribution of our proposed method.
Before details on the description of structure-based feature extraction process, we first define the structural neighborhood.
Structure-based features are defined for the amino acid at the variant site, and a cumulative value representing the amino acids in the neighborhood of the query amino acid.
The neighborhood is defined as the set of amino acid residues that lie within a specified radius of the physical structure of the protein centered around the query variant site.
Based on the relative Cartesian coordinates of the amino acids in the PDB files, we consider the neighborhood as all the residues located within the Euclidean distance of 9 Angstroms (\AA), as shown in Fig \ref{fig:neighbourhood}.
We take summaries of the feature values calculated for neighboring amino acids as the neighborhood features.
Also, the number of residues in the neighborhood (\textit{nNum}) is also included as one of the neighborhood features.
Since neighborhood information is extracted from a structure, all the neighborhood features are we considered to be structure features.

\subsubsection*{Structure-based features}
Next, we provide details on the extraction of eleven structure-based features that combine information from multiple homologous structures.
Again, the eleven features depend on the location of the variant and not on the actual change and are considered prior knowledge.
Also, unlike many existing methods, we emphasize the importance of obtaining information from multiple structures.
A PDB structure is only a static snapshot of the true protein and the functional information such as binding event may exist in one structure while not on another depending on the experimental settings, thus examining all homologous structures is important.

The first structure-based feature is the binding site information (\textit{Binding}).
The binding feature is generated by finding ligands, ions, and other biomolecules that are within a threshold distance to the protein structure.
For small molecules, i.e., ion binding site and protein-ligand binding site, we used a threshold of 4.5\AA.
For larger binding targets, i.e., DNA, RNA, and protein-protein binding sites, we used a threshold of 8\AA.
If a target molecule is observed within the threshold, then the corresponding binding site bin from the five categories is set to 1.
Binding information from multiple structures is integrated using binary \textit{OR} operation to represent all cases of possible binding sites at the query site among multiple structures.
Based on the five binding target information extracted from the structure, we add the values such that the resultant feature represents the number of different binding types found at the query site.
Binding information for the neighbourhood (\textit{nBinding}) is calculated by binary \textit{OR} the binding information from structural neighbours.

The second type of structure-based feature is hydrophobicity (\textit{KDmean}).
KDmean is the mean value of Kyle Doolittle (KD) hydrophobicity \cite{kyte1982simple} values of amino acids observed at the query location from multiple aligned structures.
The mean KD hydrophobicity value of the structural neighborhood is also used (\textit{nKD}).

The third type of features is the solvent accessibility.
Solvent accessibility is the surface area which is reachable by a solvent and is measured by squared Angstroms.
DSSP \cite{kabsch1983dictionary, touw2015series} is used to calculate the initial solvent accessibility.
After which the accessibility value is divided by the maximum exposure value possible for the amino acid triplets centered around the mutation point to obtain Residual Solvent Accessibility (RSA) value.
The maximum RSA value (\textit{RSAmax}) is chosen as a feature to represent information from multiple structures.
Mean of the maximum RSA values of amino acids in the neighborhood (\textit{nRSA}) is also used.

Fourth feature type is B factor.
B factor is a temperature dependent parameter which describes the displacement of atomic positions from their mean position that indirectly accounts for the flexibility of the locus.
This feature for a given amino acid is obtained directly from the PDB files.
We use the standard deviation of B-factor (\textit{Bstddev}) values to represent the variability of this parameter in multiple structures as a feature and the mean B-factor value of amino acids in the neighborhood (\textit{nB})as the neighbor feature.

Fifth feature type is based on sequence conservation. The sequence conservation of the locus is considered as a sequence-based property, and the method of calculating this value is described in the next section.
However, since the neighborhood is a structurally defined concept, mean of sequence conservation of amino acids in the neighborhood (\textit{nSC}) is included as a structure-based feature.

The last structure-based feature type comes from Ramachandran plot.
The Ramachandran plot depicts the phi-psi angles of amino acids in a given protein structure.
The Ramachandran plot is partitioned into regions where a secondary structure is more likely to be found.
To generate related feature we use PolyPhen2 web server to get MapReg feature for a locus.
Based on the MapReg feature generated, we represent our feature as a binary value (\textit{Mapreg}) describing whether the mutation site belongs to core regions(represented by symbols A, L, B, P), or not.

\subsection*{Extracting sequence-based features}
There are five sequence-based features tested that are specific to the prior loci-based information about the variant and are not dependent on the change.
The first is the sequence conservation (\textit{SeqCons}).
The sequence conservation implies the how conserved the locus is throughout the evolution process.
The assumption being that the locus that is unchanged over the evolution process must be functionally important.
The sequence conservation is indirectly computed by calculating the variation of amino acids in loci.
That is, after multiple sequences aligning via ClustalW \cite{larkin2007clustal, goujon2010new} the query sequence with the homologous sequence extracted from MetaPhOrs API \cite{pryszcz2010metaphors}, a variation of an amino acid at a mutation point is calculated and conservation score is computed using the Jensen-Shannon divergence.
Four additional features are adapted from sequence-based features of PolyPhen2 from DNA and protein levels.
At the DNA level, the position variant locus within the codon (\textit{CodPos}) and the distance of the variant locus from the closest intron/exon junction (\textit{MinDJxn}) are identified and included as a sequence-based features.
At the protein level, Position-Specific Independent Counts (\textit{PSIC}) score \cite{sunyaev1999psic} of the wild type amino acid residue and the number of amino acids observed at variant locus (\textit{Nobs}) during multiple sequence alignment of homologous sequences are used.
The list of features is summarized in Tab~\ref{table:list-of-features}.

\begin{table}[h]
 \centering
 \caption{List of structure- and sequence-based features. The rank represents the importance of individual features obtained by 5-fold cross validation on the HumDiv training data using MLP for attribute evaluation.}
 \label{table:list-of-features}
 \begin{tabular}{p{0.15\textwidth}|p{0.65\textwidth}|p{0.05\textwidth}}
  \hline
  Name & Description & Rank\\
  \hline
  \multicolumn{3}{l}{\textbf{Structure based features}}\\
  \hline
  nNum & number of amino acids whin structural neighborhood & 2\\
  RSAmax & maximum residual solvent accessibility value& 3\\
  nRSA & mean residual solvent accessibility of structural neighbors & 4 \\
  nB & mean B factor of structural neighbors & 5\\
  nBinding & whether neighbors have a binding site or not& 7\\
  nKD & mean KD hydrophobicity of structural neighbors & 9\\
  Bstddev & standard deviation of B factors & 10\\
  Mapreg & whether variant locus is within core region of phi-psi Ramachandran map (PolyPhen)& 11\\
  nSC & mean sequence conservation of structural neighbors & 12\\
  KDmean & mean KD hydrophobicity value & 13\\
  Binding & whether site is binding or not & 14\\
  \hline
  \multicolumn{3}{l}{\textbf{Sequence based features}}\\
  \hline
  PSIC & PSIC score of wild type amino acid (PolyPhen)& 1\\
  SeqCons & sequence conservation & 6 \\
  Nobs & number of amino acid observed at the substitution position in the multiple alignment (PolyPhen) & 8\\
  CodPos & position of variant locus within a codon (PolyPhen) & 15\\
  MinDJxn & distance of variant locus from closest exon/ intron junction (PolyPhen)& 16\\
  \hline
 \end{tabular}
\end{table}

\section*{Results}
\subsection*{Classification performance}
Classification accuracies of seven machine learning algorithms are measured to validate the effectiveness of the proposed loci-based features.
The goal of the learning task is to classify a variant site as pathogenic or benign based on the features defined at the variant locus.
For HumDiv dataset, the data is randomly split such that 80\% is used for training and validation and 20\% is used for testing.
We additionally test the model on ClinVar dataset, i.e. small but newer and cleaner dataset, using the learned model.
The 20\% of training data is used as a validation set to tune hyperparameter of each machine learning algorithms.
Seven machine learning algorithms used are naive Bayes, support vector machine (SVM; LibSVM implementation \cite{chang2011libsvm}), logistic regression, multi-layer perceptron (MLP), k-nearest neighbors(KNN), decision table, and random forest.
All experiments are performed using Weka \cite{hall2009weka, witten2016data} and Weka's unsupervised attribute filter is used for replacing missing values in the test set based on means and modes of training data.

To validate the performance of the features, we use weighted F-score calculated by Weka.
The precision-recall based scores are known to be a measurement of choice when there is a class imbalance as in the case of our test sets.
The weighted F score is a weighted average of the precision and recall that account for both the correctly predicted positives as well as incorrectly predicted positive and negatives. The weighted F score reaches its best value at 1 and worst at 0.

\subsubsection*{Performance variation of learning methods}
First, we test the models learned on HumDiv test datasets consisting of 1094 SNVs.
The summary of the weighted F score values are shown in Tab~\ref{table-algo-results}.
We can see that random forest provide the best scores going up as high as weighted F score of 0.85.

\begin{table}[h]
 \centering
 \caption{ {\bf Classification performance on HumDiv dataset.} Summary of weighted F scores for each algorithm using sequence-based features, structure features and the combination of sequence and structure features (combined).}
 \label{table-algo-results}
 \begin{tabular}{cccc}
  \hline
  Algorithm & Sequence only & Structure only & Combined\\
  \hline
  Naive Bayes & 0.82 & 0.71 & 0.80 \\
  SVM & 0.79 & 0.69 & 0.81 \\
  Logistic regression & 0.79 & 0.69 & 0.80 \\
  \textbf{MLP} & 0.78 & 0.71 & 0.82 \\
  KNN & \textbf{0.83} & 0.72 & 0.80 \\
  Decision table & 0.81 & 0.74 & 0.82 \\
  \textbf{Random forest} & \textbf{0.83} & \textbf{0.76} & \textbf{0.85} \\
  \hline
 \end{tabular}

\end{table}

Then we test if the model can be transferred to test the ClinVar dataset consisting of 438 loci with 286 pathogenic and 152 benign labels in 101 genes.
The Tab~\ref{table-algo-results2} summaries the results.
Although MLP does have the highest F score in the ClinVar test case, there is no clear winner in terms of learning methods.
Also, compared to the result of HumDiv test case, F score decreased overall with highest weighted F score being 0.75.
This was expected since ClinVar labeling and HumDiv labeling criterion and the selection of benign cases are slightly different.
With a consistent dataset, we expect that the F score will increase.
With the difference in the distribution of training set and testing set in mind, we will focus our analysis using ClinVar dataset which consists of labels that are better curated.
Also, the choice of learning method was unimportant and we were able to get stable results using most of the learning methods.

\begin{table}[h]
 \centering
 \caption{ {\bf Classification performance on ClinVar dataset.} Summary of weighted F scores for each algorithm using sequence-based features (sequence), structure-based features (structure) and the combination of sequence- and structure-based features (combined).}
 \label{table-algo-results2}
 \begin{tabular}{cccc}
  \hline
  Algorithm & Sequence & Structure & Combined\\
  \hline
  Naive Bayes & 0.69 & 0.69 & 0.72 \\
  SVM & 0.70 & 0.66 & 0.72 \\
  Logistic regression & 0.70 & 0.65 & 0.72 \\
  \textbf{MLP} & \textbf{0.70} & 0.68 & \textbf{0.75} \\
  KNN & 0.69 & 0.70 & 0.71\\
  Decision table & 0.68 & 0.64 & 0.70 \\
  \textbf{Random forest} & 0.69 & \textbf{0.69} & 0.72 \\
  \hline
 \end{tabular}
\end{table}

\subsubsection*{Performance comparison between sequence- and structure-based features}
Tested on HumDiv dataset, the sequence-based features, with a maximum value of 0.83, performed better than structure-based features, with a maximum value of 0.76.
Interestingly the difference was minor for the ClinVar dataset with maximum weighted F score of 0.70 for sequence-based and 0.69 for structure-based features.
The weighted F score for combined features performed best for the majority of the tested learning methods in HumDiv test dataset and all learning methods in ClinVar dataset.
This shows that, although the performance of structure-based features is similar or even lower, structure-based features are contributing in terms of predicting the important of the loci in disease.
We again emphasize that the current features used are prior information about the importance of a locus in the disease process.

In addition to weighted F score comparison, we examined the importance of individual features on the training data using MLP.
The feature ranks are shown in the third column of Tab~\ref{table:list-of-features}.
Four structure-based features and one sequence-based features are ranked within top 5 and three out of four structure-based features where neighborhood features.
Overall, the results tell us that structure-based features are informative and that structure-based features and sequence-based features are complementary.

\subsubsection*{Importance of neighborhood information}
We next tested the contribution of structural neighborhood features in improving the classification performance.
To do this, the performance of structure-based features with and without the neighborhood information was measured as shown in Tab~\ref{table-neighimp-results2}.
It can be seen that neighborhood information improves the weighted F score for learning algorithms tested on HumDiv test and ClinVar datasets.

\begin{table}[h]
 \centering
 \caption{Classification performance of structure-based features with and without neighbourhood information. The values represent the weighted F scores.}
 \label{table-neighimp-results}
 \begin{tabular}{cccc}
  \hline
  Dataset & Algorithm & Without Neighbourhood & With Neighbourhood \\
  \hline
  HumDiv & MLP & 0.70 & \textbf{0.71}  \\
  HumDiv & Random Forest & 0.73 & \textbf{0.76} \\
        ClinVar & MLP & 0.64 & \textbf{0.68} \\
  ClinVar & Random Forest & 0.66 & \textbf{0.69} \\
  \hline
 \end{tabular}
\end{table}

\subsubsection*{Performance comparison}
We compare our selection of features that consist only of variant loci based, i.e. prior information, with the PolyPhen2 that considers both loci based features and nucleotide change based features, i.e, requiring information about the what the nucleotide change.
The ClinVar dataset is used since the PolyPhen2's models are trained on HumDiv dataset.
We compared the predictions of our features using MLP and Random Forest with PolyPhen2 predictions that are obtained from PolyPhen2 web server (URL: http://genetics.bwh.harvard.edu/pph2/).
Test results are summarized in Tab~\ref{table-compare-PolyPhen}.

Interestingly, our loci-based features performed comparable and even slightly better (MLP weighted F score of 0.75 compared to 0.74) than the PolyPhen2 predictions that use nucleotide change based information in addition to loci-based information.
To test how much the nucleotide change information contributes, we also tested PolyPhen2's six nucleotide change (DNA change) based features, as listed in the Supplementary text: \nameref{DNA_change}, in addition to our loci-based features.
With the added features the weighted F scores improved as shown in third column of Tab~\ref{table-compare-PolyPhen}.
We emphasize again that the purpose of the work is to test how much a non-statistics based prior information contributes in the determination of the pathogenic effect of variant such that it can be used toward rare variant rare disease cases.
With the purpose in mind, interestingly variant loci seem to contribute largely in determining the pathogenic effects.

\begin{table}[h]
 \centering
 \caption{Comparison of algorithm performance with PolyPhen2 predictions. The values represent the weighted F scores for ClinVar dataset.}
 \label{table-compare-PolyPhen}
 \begin{tabular}{|c|c|c|c|}
  \hline
  \multicolumn{3}{|c|}{Our proposed features via MLP} & PolyPhen2 \\
  \hline
  Algorithm & Loci based & DNA change & \multirow{3}{*}{0.74}\\
  \cline{1-3}
  MLP & \textbf{0.75} & \textbf{0.79} &  \\
  \cline{1-3}
  Random Forest & 0.72 & \textbf{0.78} &  \\
  \hline
 \end{tabular}
\end{table}

\subsection*{Examining rare variants in rare disease cases}
To analyze the performance of the model for the rare variant in rare disease cases which is the goal of the work, we identify rare variants (MAF$<$0.01) in the dataset and check if they were related to a rare disease.
The ClinVar dataset contains 163 pathogenic rare variants and 117 benign rare variants as described in the dataset section.

The Tab~\ref{table-RV-RD} summarizes weighted F score for sequence-based (sequence), structure-based (structure), sequence- and structure-based features combined (combined), and nucleotide change included features (DNA change).
Comparing the number of correctly predicted benign cases (52 vs 79) and pathogenic cases (137 vs 110) we can see that sequence-based features were able to predict more pathogenic cases and structure-based features were able to predict benign cases better.
Combined features increased the number of correctly predicted and adding DNA change information improved the result slightly more by increasing the number of correctly predicted pathogenic cases.

\begin{table}[h]
 \centering
 \caption{Rare variant and rare disease SNVs predictions, based on prior information. The values represent the number of correctly classified (out of 117 benign, 163 pathogenic) based on MLP prediction model.}
 \label{table-RV-RD}
 \begin{tabular}{|c|c|c|c|c|c|}
  \hline
  & \multicolumn{4}{c|}{Our proposed features via MLP} & \multirow{2}{*}{PolyPhen2} \\
  \cline{2-5}
  & Sequence & Structure & Combined & DNA change&\\
  \hline
  Benign (TN) & 52 & \textbf{79} & 68 & 62 & 49\\
  Pathogenic (TP) & 137 & 110 & 137 & \textbf{149} & \textbf{150} \\
  \hline TP+TN & 189 & 189 & 205 & \textbf{211} & 199 \\
  \hline
  weighted F score & 0.66 & 0.68 & \textbf{0.73} & \textbf{0.74} & 0.69  \\
  \hline
 \end{tabular}
\end{table}

\subsubsection*{Structural examination of rare variant in rare disease}
To understand how the structure-based features perform in more detail, we take a detailed look at cases where structure-based features performed well while all others were not successful.

\begin{figure}[h!]
 \centering
 \includegraphics[width=.8\textwidth]{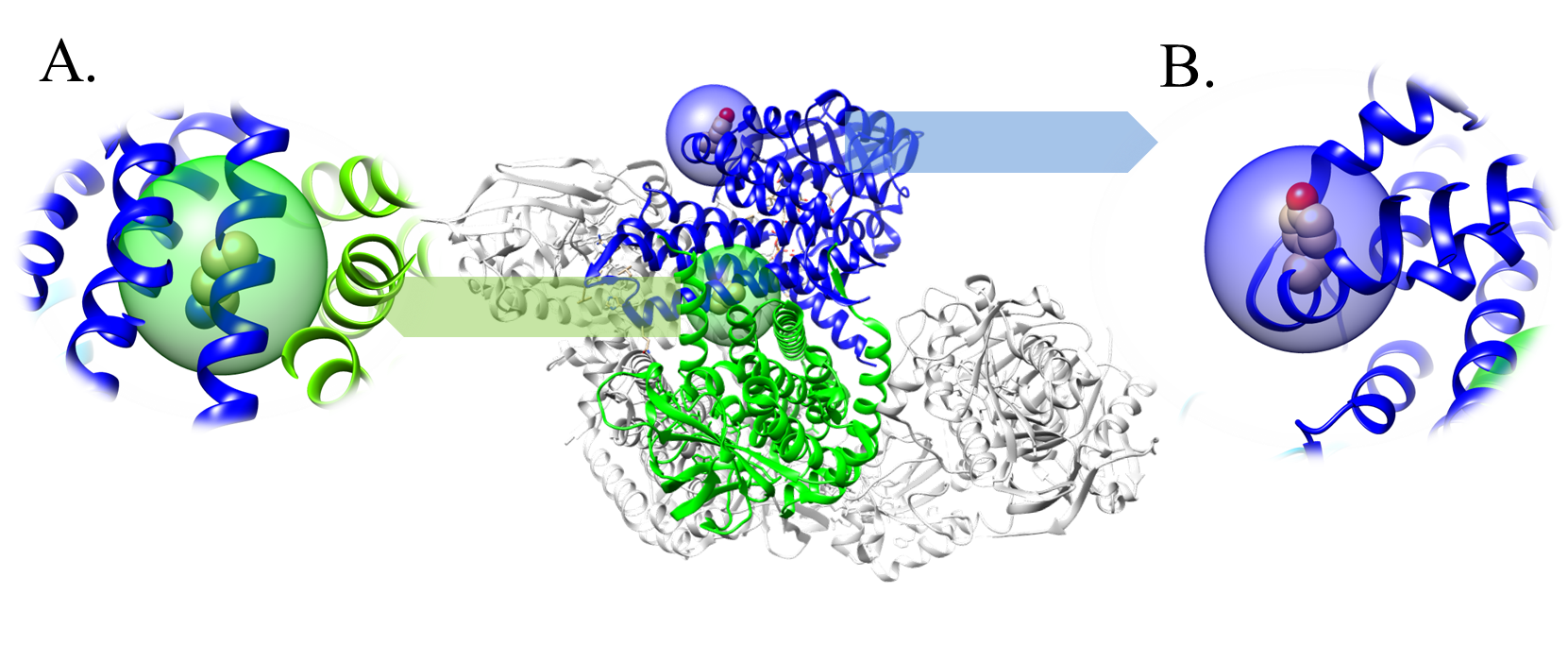}
 \caption{\textbf{Two pathogenic rare variants ACADM gene}
  The SNV loci of  Chr1:75761161 and Chr1:75732724 corresponding to ACADM gene mapped to PDBID:2A1T.A positions 304 in \textbf{A} (green) and 42 in \textbf{B} (blue), respectively.
       An SVN locus depicted in spheres and neighborhoods in transparent spheres. The two variant loci are known to be pathogenic in Medium-chain acyl-coenzyme A dehydrogenase deficiency. }
 \label{fig:rv1}
\end{figure}

First, we examine two rare variants in chromosome 1 positions 75761161 and 75732724.
The two loci are associated with Medium-chain acyl-coenzyme A dehydrogenase deficiency (MCADD) which is a metabolic disorder that prevents the body from converting certain fats to energy and variants in ACADM gene is known to cause this MCADD in an autosomal recessive manner.
The SNV locus Chr1:75761161 depicted in Fig~\ref{fig:rv1}\textbf{A} shows that the locus is a domain binding site.
The structure-based features also shows that the neighborhood is mostly hydropathic (nKD=1.14) and is surrounded by many residues (nNum=17).
Also, SNV locus Chr1:75732724 depicted in Fig~\ref{fig:rv1}\textbf{B} shows that the locus is a surface region and a not binding site, however, since the neighborhood is hydropathic and stable (relatively low nB of 38), the locus may be part of a transient binding site.
For both loci, the sequence base features predict this SNV locus as benign with no strong features to indicate pathogenic effects.

\begin{figure}[h!]
 \centering
 \includegraphics[width=.8\textwidth]{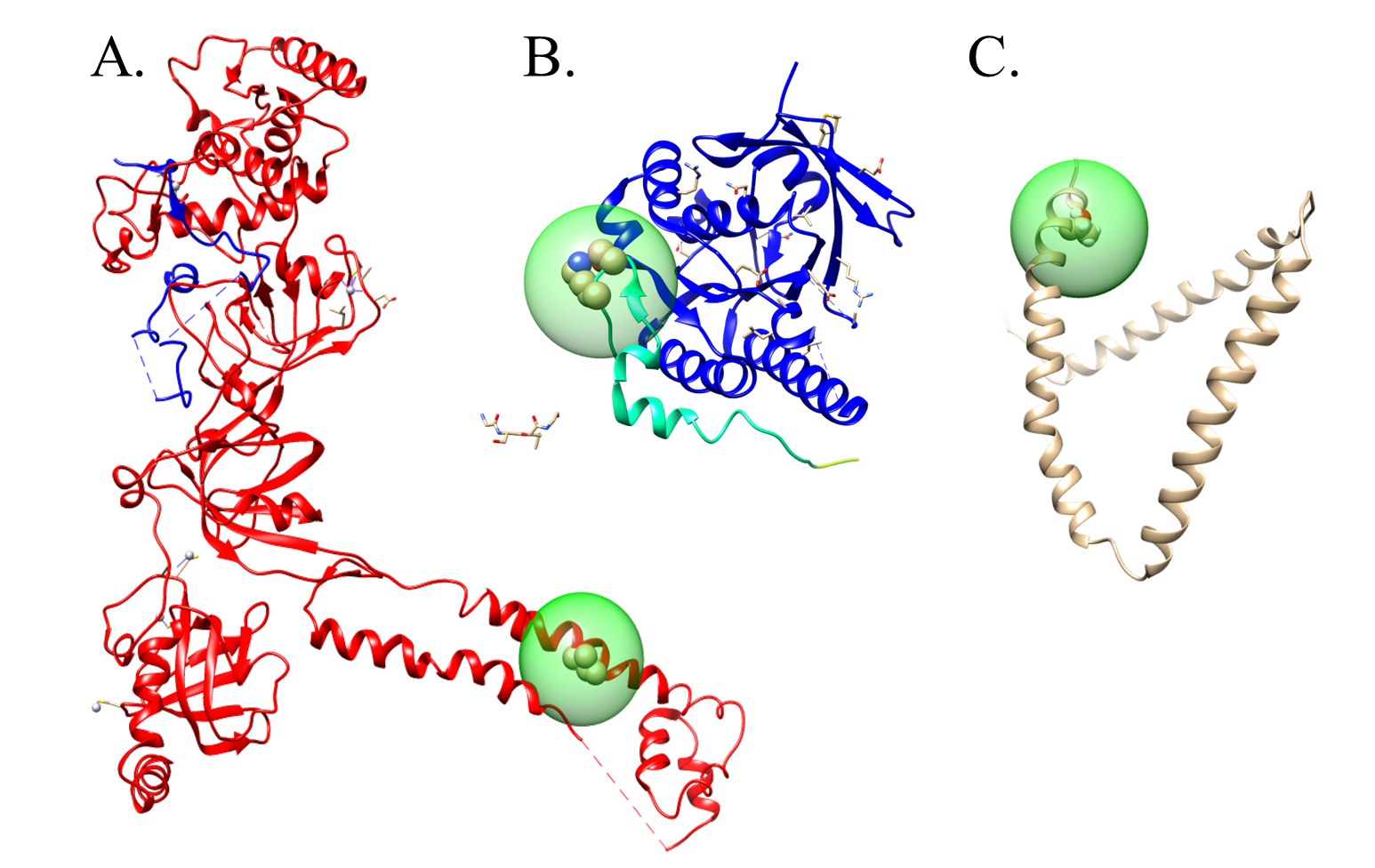}
 \caption{\textbf{Three benign rare variants in the flexible solvent accessible areas}
    A. Shows SNV locus Chr13:32379392 of BRCA2 gene mapped to PDBID:1MIU.A position 2865.
    B. Shows SNV locus Chr13:32338940 ($NM_000059.3:c.8567A>C GRCh38$) of BRCA2 gene mapped to PDBID:1N0W.B position 1529.
    C. Shows SNV locus Chr21:34370500 of KCNE2 gene mapped to PDBID:2M0Q.A position 8.
   Variant loci are depicted in spheres and neighborhoods in transparent spheres.
   All three SNV loci are known to benign. }
 \label{fig:rv2}
\end{figure}

We next look at three SNV loci that are located at flexible solvent accessible areas.
Fig~\ref{fig:rv2}\textbf{A} and \ref{fig:rv2}\textbf{B} show two protein structures that are associated with BRCA2 gene. The SNVs at the two loci are known to be benign.
Looking at SNV locus Chr13:32379392 in Fig~\ref{fig:rv2}\textbf{A}, we can see that it is part of tower region of the protein structure (PDBID:1N0W) with high exposure to solvent (nRSA=0.46) with low hydrophobicity value (nKD=-1.03) which says that it is not directly binding region although the three helical bundle at the tip of the tower is known to bind to DNA \cite{Yang2002}.
SNV locus in Fig~\ref{fig:rv2}\textbf{B} is part of BRC repeat 4 (BRC4) that is bound to RAD51.
BRC repeats are known to be well conserved and are the primary sites through which BRCA2 binds to RAD51 that has a critical role in DNA recombination \cite{Pellegrini2002}.
In contrary to the suggestion of the structure provider of PDBID:1N0W that the variant locus should be pathogenic \cite{Pellegrini2002}, ClinVar's expert panel find the variant benign.
We find that this region, although it is a binding site, is beta turn region which is more flexible solvent accessible region.
Fig~\ref{fig:rv2}\textbf{C} depicts SNV locus Chr21:34370500 of KCNE2 gene mapped to PDBID:2M0Q.A at position 8. 2M0Q is a transmembrane helix structure and the SNV locus is positioned at the tip of a more flexible area of the structure.
With the three cases, we find that variant at loci of flexible solvent accessible areas tends to be benign.

\begin{figure}[h!]
 \centering
 \includegraphics[width=0.8\textwidth]{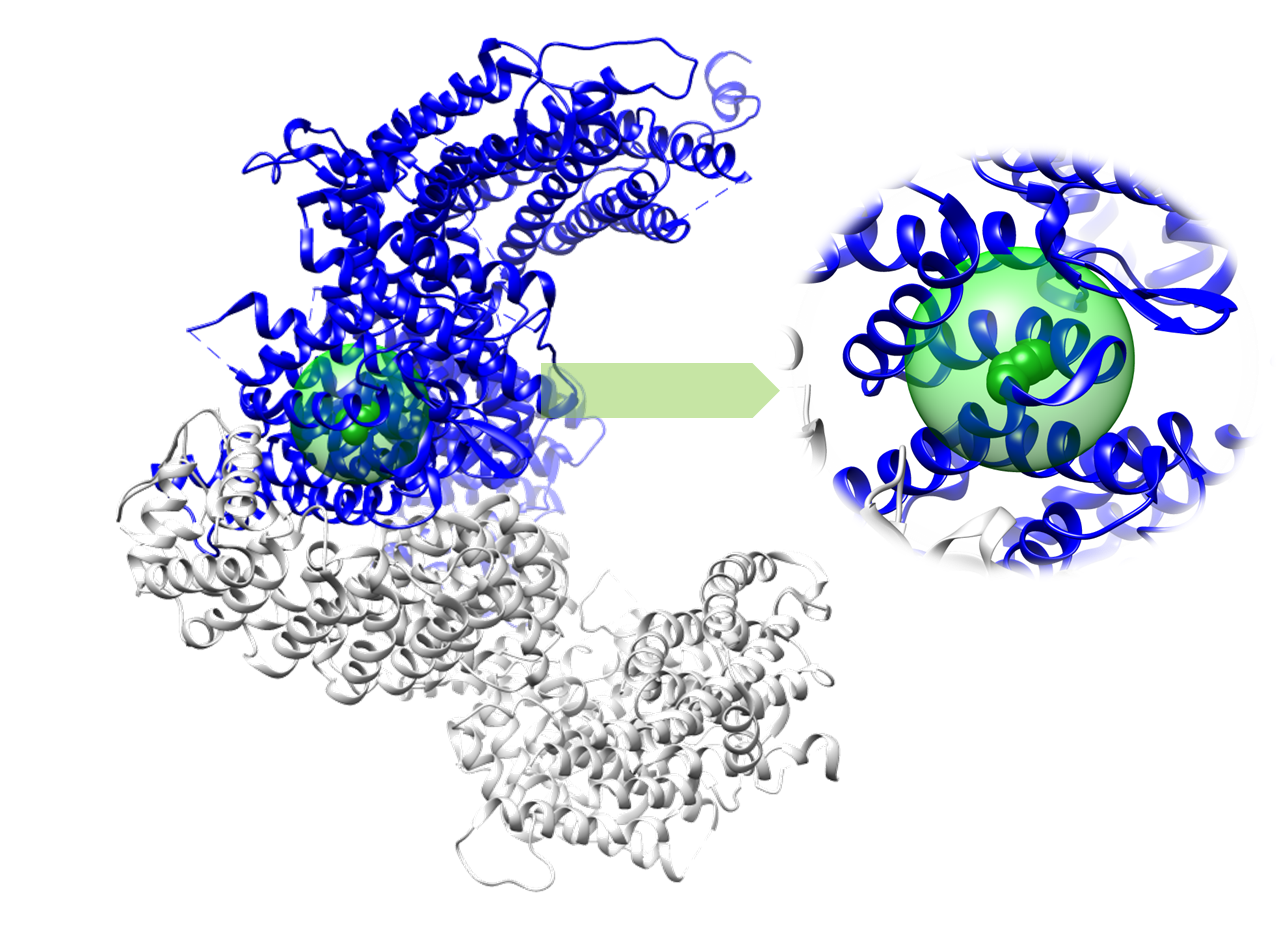}
 \caption{\textbf{A benign rare variant in FANCI gene}
  The SNV locus of Chr15:89285210 corresponding to FANCI gene map to PDBID:3S4W.A positions 604.
        Variant loci is depicted in spheres and neighborhoods in transparent spheres.
        The two variant loci are known to be benign in Fanconi anemia. }
 \label{fig:rv3} 
\end{figure}

Final we look a benign SNV locus Chr15:89285210 corresponding to FANCI gene.
FANCI gene is known to be associated with Fanconi anemia which is a rare disease in the category of bone marrow failure syndromes.
Fanconi anemia is often associated with a progressive deficiency of all bone marrow production of blood cells, red blood cells, white blood cells, and platelets.
Fig~\ref{fig:rv3} depicts PDBID:3S4W which FANCI structure (blue) is bound to FANCD2 (gray) structure. The two complex is known to bind to DNA structure at the center of the curved region.
The SNV locus is completely buried but has relatively flexible with high B-score of 70.
The region is not associated with protein-protein binding nor protein-DNA binding.
Structure-based feature predicts this as benign while as sequence-based features and HumDiv predicts it pathogenic.

\section*{Discussions and Conclusions}
Determination of whether a variant or group of variants have pathogenic effects involves various factors.
To understand the effect of variants better we separate out the features that can be predetermined based on the locus of the variant and the features that are dependent on the changed in the DNA sequence.
The loci-based information can be considered as prior information since the importance of loci can be pre-calculated.
Also, structure-based loci features do not depend on statistical means which makes it attractive for rare variant rare disease cases.

Our results show that sequence-based features perform slightly better than structure-based features in classify pathogenic variants.
However, structure-based features provide a complementary information and when combined with sequence-based features, it improves the F score up to 0.85 in the HumDiv dataset.
In theory, the sequence contains all necessary information, however, we still don't know how and structure-based information can fill in this knowledge gap.

In terms of structure-based features, there are two novel contributions that seem to improve the performance: we use multiple homologous structures and we used features from the structural neighborhood.
The newly introduce structure-based features in combination with sequence-based features of loci information only performs comparably to the PolyPhen2 that utilized loci information as well as the DNA change information.
Also, empirical structure-based features of rare variant in rare disease cases show that structure-based features play an important role in correctly classifying benign verse pathogenic cases.

Many previous causal variant prediction methods consider structure-based features as minor factors even though there are several works that say structure-based features are important.
It is true that structures are more difficult to work with compared to sequences and extracting structure-based features from given structures are not trivial.
Even after successfully extracting features from structures, correctly mapping it back to the original DNA sequence is difficult due to several difficulties, including differential splicing and miss match in a start position of reference protein sequence to the PDB protein sequence is challenging.
Also, structure-based features have relatively small coverage and data available compared to sequential data.
Even with this challenges, our work has shown that loci of the variant event plays a key role in determining the pathogenic effect and showed that the structure-based features can in cases classify better as shown in rare variant rare disease cases.
In the future, we would like to include more structure-based features utilizing computational methods such as binding site predictions, and stability predictions and also increase coverage by including structure prediction methods.

\section*{Supporting information}


\paragraph*{S1 Table.} %
\label{DNA_change}
{\textbf{Nucleotide change based features.} To identify the information gain, we also tested six variant-based features in addition to loci information for evaluation.
Each of features are obtained from PolyPhen2 server (URL: http://genetics.bwh.harvard.edu/pph2/dokuwiki/appendix\_a
).} \\
\begin{tabular}{l|p{0.8\textwidth}}
\hline
Feature & Description\\
\hline
 Score2 & the PSIC score for mutant amino acid \\
 dScore & represents the difference \textit{PSIC} and \textit{Score2} \\
 Transv & represents if the substitution is a transversion\\
 CpG & values represent change in CpG content due to substitution\\
 IdPmax & represents mutant amino acid's maximum congruency in multiple sequence alignment\\
 IdQmin & represents the sequence identity to closest homologue deviating from wild type amino acid\\
\hline
\end{tabular}

\paragraph*{S2 HumDiv training set.} %
\label{HumDiv_train}
{
List of HumDiv training set features and meta data.
}

\paragraph*{S3 HumDiv test set.} %
\label{HumDiv_test}
{
List of HumDiv test set features and meta data.
}
\paragraph*{S4 ClinVar data set.} %
\label{ClinVar_test}
{
List of ClinVar test set features and meta data.
}

\bibliographystyle{plain}

\bibliography{SNP_prioritization}

\begin{thebibliography}{10}

\bibitem{adzhubei2013predicting}
Ivan Adzhubei, Daniel~M Jordan, and Shamil~R Sunyaev.
\newblock Predicting functional effect of human missense mutations using
  polyphen-2.
\newblock {\em Current protocols in human genetics}, pages 7--20, 2013.

\bibitem{adzhubei2010method}
Ivan~A Adzhubei, Steffen Schmidt, Leonid Peshkin, Vasily~E Ramensky, Anna
  Gerasimova, Peer Bork, Alexey~S Kondrashov, and Shamil~R Sunyaev.
\newblock A method and server for predicting damaging missense mutations.
\newblock {\em Nature methods}, 7(4):248--249, 2010.

\bibitem{aken2016ensembl}
Bronwen~L Aken, Sarah Ayling, Daniel Barrell, Laura Clarke, Valery Curwen,
  Susan Fairley, Julio~Fernandez Banet, Konstantinos Billis, Carlos~Garc{\'\i}a
  Gir{\'o}n, Thibaut Hourlier, et~al.
\newblock The ensembl gene annotation system.
\newblock {\em Database}, 2016:baw093, 2016.

\bibitem{altschul1990basic}
Stephen~F Altschul, Warren Gish, Webb Miller, Eugene~W Myers, and David~J
  Lipman.
\newblock Basic local alignment search tool.
\newblock {\em Journal of molecular biology}, 215(3):403--410, 1990.

\bibitem{bamshad2011exome}
Michael~J Bamshad, Sarah~B Ng, Abigail~W Bigham, Holly~K Tabor, Mary~J Emond,
  Deborah~A Nickerson, and Jay Shendure.
\newblock Exome sequencing as a tool for mendelian disease gene discovery.
\newblock {\em Nature reviews. Genetics}, 12(11):745, 2011.

\bibitem{berman2003announcing}
Helen Berman, Kim Henrick, and Haruki Nakamura.
\newblock Announcing the worldwide protein data bank.
\newblock {\em Nature Structural \& Molecular Biology}, 10(12):980--980, 2003.

\bibitem{Boycott2013}
Kym~M. Boycott, Megan~R. Vanstone, Dennis~E. Bulman, and Alex~E. MacKenzie.
\newblock {Rare-disease genetics in the era of next-generation sequencing:
  discovery to translation}.
\newblock {\em Nature Reviews Genetics}, 14(10):681--691, 2013.

\bibitem{Bullock2000}
a~N Bullock, J~Henckel, and a~R Fersht.
\newblock {Quantitative analysis of residual folding and DNA binding in mutant
  p53 core domain: definition of mutant states for rescue in cancer therapy.}
\newblock {\em Oncogene}, 19(10):1245--1256, mar 2000.

\bibitem{Carter2009}
Hannah Carter, Sining Chen, Leyla Isik, Svitlana Tyekucheva, Victor~E
  Velculescu, Kenneth~W Kinzler, Bert Vogelstein, Rachel Karchin, and Feature
  Selection.
\newblock {Cancer-specific high-throughput annotation of somatic mutations:
  computational prediction of driver missense mutations.}
\newblock {\em Cancer Research}, 69(16):6660--6667, aug 2009.

\bibitem{chang2011libsvm}
Chih-Chung Chang and Chih-Jen Lin.
\newblock Libsvm: a library for support vector machines.
\newblock {\em ACM Transactions on Intelligent Systems and Technology (TIST)},
  2(3):27, 2011.

\bibitem{chial2008mendelian}
Heidi Chial.
\newblock Mendelian genetics: patterns of inheritance and single-gene
  disorders.
\newblock {\em Nature Education}, 1(1):63, 2008.

\bibitem{de2013background}
R~de~Vrueh, ERF Baekelandt, and JMH de~Haan.
\newblock Background paper 6.19 rare diseases.
\newblock Technical report, World Health Organization, Geneva, 2013.

\bibitem{durinck2005biomart}
Steffen Durinck, Yves Moreau, Arek Kasprzyk, Sean Davis, Bart De~Moor, Alvis
  Brazma, and Wolfgang Huber.
\newblock Biomart and bioconductor: a powerful link between biological
  databases and microarray data analysis.
\newblock {\em Bioinformatics}, 21(16):3439--3440, 2005.

\bibitem{freed2012mutant}
William~A Freed-Pastor and Carol Prives.
\newblock Mutant p53: one name, many proteins.
\newblock {\em Genes \& development}, 26(12):1268--1286, 2012.

\bibitem{GlobalGenes}
{Global Genes}.
\newblock {RARE Diseases: Facts and Statistics}, 2017.

\bibitem{goujon2010new}
Mickael Goujon, Hamish McWilliam, Weizhong Li, Franck Valentin, Silvano
  Squizzato, Juri Paern, and Rodrigo Lopez.
\newblock A new bioinformatics analysis tools framework at embl--ebi.
\newblock {\em Nucleic acids research}, 38(suppl 2):W695--W699, 2010.

\bibitem{Gress2016}
Alexander Gress, Vasily Ramensky, Joachim B{\"{u}}ch, Andreas Keller, and
  Olga~V Kalinina.
\newblock {StructMAn: annotation of single-nucleotide polymorphisms in the
  structural context.}
\newblock {\em Nucleic acids research}, 44(18):W463–W468, 2016.

\bibitem{hall2009weka}
Mark Hall, Eibe Frank, Geoffrey Holmes, Bernhard Pfahringer, Peter Reutemann,
  and Ian~H Witten.
\newblock The weka data mining software: an update.
\newblock {\em ACM SIGKDD explorations newsletter}, 11(1):10--18, 2009.

\bibitem{jiang2007sequence}
Rui Jiang, Hua Yang, Linqi Zhou, C-C~Jay Kuo, Fengzhu Sun, and Ting Chen.
\newblock Sequence-based prioritization of nonsynonymous single-nucleotide
  polymorphisms for the study of disease mutations.
\newblock {\em The American Journal of Human Genetics}, 81(2):346--360, 2007.

\bibitem{Joerger2006}
Andreas~C Joerger, Hwee~Ching Ang, and Alan~R Fersht.
\newblock {Structural basis for understanding oncogenic p53 mutations and
  designing rescue drugs.}
\newblock {\em Proceedings of the National Academy of Sciences of the United
  States of America}, 103(41):15056--61, oct 2006.

\bibitem{kabsch1983dictionary}
Wolfgang Kabsch and Christian Sander.
\newblock Dictionary of protein secondary structure: pattern recognition of
  hydrogen-bonded and geometrical features.
\newblock {\em Biopolymers}, 22(12):2577--2637, 1983.

\bibitem{kapustin2008splign}
Yuri Kapustin, Alexander Souvorov, Tatiana Tatusova, and David Lipman.
\newblock Splign: algorithms for computing spliced alignments with
  identification of paralogs.
\newblock {\em Biology direct}, 3(1):20, 2008.

\bibitem{kircher2014general}
Martin Kircher, Daniela~M Witten, Preti Jain, Brian~J O'roak, Gregory~M Cooper,
  and Jay Shendure.
\newblock A general framework for estimating the relative pathogenicity of
  human genetic variants.
\newblock {\em Nature genetics}, 46(3):310--315, 2014.

\bibitem{kyte1982simple}
Jack Kyte and Russell~F Doolittle.
\newblock A simple method for displaying the hydropathic character of a
  protein.
\newblock {\em Journal of molecular biology}, 157(1):105--132, 1982.

\bibitem{landrum2016clinvar}
Melissa~J Landrum, Jennifer~M Lee, Mark Benson, Garth Brown, Chen Chao,
  Shanmuga Chitipiralla, Baoshan Gu, Jennifer Hart, Douglas Hoffman, Jeffrey
  Hoover, et~al.
\newblock Clinvar: public archive of interpretations of clinically relevant
  variants.
\newblock {\em Nucleic acids research}, 44(D1):D862--D868, 2016.

\bibitem{larkin2007clustal}
Mark~A Larkin, Gordon Blackshields, NP~Brown, R~Chenna, Paul~A McGettigan,
  Hamish McWilliam, Franck Valentin, Iain~M Wallace, Andreas Wilm, Rodrigo
  Lopez, et~al.
\newblock Clustal w and clustal x version 2.0.
\newblock {\em bioinformatics}, 23(21):2947--2948, 2007.

\bibitem{Lo2015}
Yu~Shu Lo, Sing~Han Huang, Yong~Chun Luo, Chun~Yu Lin, and Jinn~Moon Yang.
\newblock {Reconstructing genome-wide protein-protein interaction networks
  using multiple strategies with homologous mapping}.
\newblock {\em PLoS ONE}, 10(1), 2015.

\bibitem{Lwin2007}
Thu~Zar Lwin, Jason~J Durant, and Donald Bashford.
\newblock {A fluid salt-bridging cluster and the stabilization of p53.}
\newblock {\em Journal of Molecular Biology}, 373(5):1334--1347, nov 2007.

\bibitem{mathe2006computational}
Ewy Mathe, Magali Olivier, Shunsuke Kato, Chikashi Ishioka, Pierre Hainaut, and
  Sean~V Tavtigian.
\newblock Computational approaches for predicting the biological effect of p53
  missense mutations: a comparison of three sequence analysis based methods.
\newblock {\em Nucleic acids research}, 34(5):1317--1325, 2006.

\bibitem{mclaren2016ensembl}
William McLaren, Laurent Gil, Sarah~E Hunt, Harpreet~Singh Riat, Graham~RS
  Ritchie, Anja Thormann, Paul Flicek, and Fiona Cunningham.
\newblock The ensembl variant effect predictor.
\newblock {\em Genome biology}, 17(1):122, 2016.

\bibitem{mueller2015ball}
Sabine~C Mueller, Christina Backes, Olga~V Kalinina, Benjamin Meder, Daniel
  St{\"o}ckel, Hans-Peter Lenhof, Eckart Meese, and Andreas Keller.
\newblock Ball-snp: combining genetic and structural information to identify
  candidate non-synonymous single nucleotide polymorphisms.
\newblock {\em Genome medicine}, 7(1):65, 2015.

\bibitem{ng2003sift}
Pauline~C Ng and Steven Henikoff.
\newblock Sift: Predicting amino acid changes that affect protein function.
\newblock {\em Nucleic acids research}, 31(13):3812--3814, 2003.

\bibitem{Ng2003}
Pauline~C. Ng and Steven Henikoff.
\newblock {SIFT: predicting amino acid changes that affect protein function}.
\newblock {\em Nucleic Acids Research}, 31(13):3812--3814, jul 2003.

\bibitem{ng2010exome}
Sarah~B Ng, Kati~J Buckingham, Choli Lee, Abigail~W Bigham, Holly~K Tabor,
  Karin~M Dent, Chad~D Huff, Paul~T Shannon, Ethylin~Wang Jabs, Deborah~A
  Nickerson, et~al.
\newblock Exome sequencing identifies the cause of a mendelian disorder.
\newblock {\em Nature genetics}, 42(1):30, 2010.

\bibitem{pearson1988improved}
William~R Pearson and David~J Lipman.
\newblock Improved tools for biological sequence comparison.
\newblock {\em Proceedings of the National Academy of Sciences},
  85(8):2444--2448, 1988.

\bibitem{Pellegrini2002}
Luca Pellegrini, David~S Yu, Thomas Lo, Shubha Anand, MiYoung Lee, Tom~L
  Blundell, and Ashok~R Venkitaraman.
\newblock {Insights into DNA recombination from the structure of a RAD51-BRCA2
  complex.}
\newblock {\em Nature}, 420(6913):287--293, 2002.

\bibitem{Pi2013}
Jaewoo Pi and Lee Sael.
\newblock {Mass spectrometry coupled experiments and protein structure modeling
  methods.}
\newblock {\em International Journal of Molecular Sciences},
  14(10):20635--20657, jan 2013.

\bibitem{pryszcz2010metaphors}
Leszek~P Pryszcz, Jaime Huerta-Cepas, and Toni Gabald{\'o}n.
\newblock Metaphors: orthology and paralogy predictions from multiple
  phylogenetic evidence using a consistency-based confidence score.
\newblock {\em Nucleic acids research}, page gkq953, 2010.

\bibitem{quesada2012exome}
V{\'\i}ctor Quesada, Laura Conde, Neus Villamor, Gonzalo~R Ord{\'o}{\~n}ez,
  Pedro Jares, Laia Bassaganyas, Andrew~J Ramsay, S{\'\i}lvia Be{\`a}, Magda
  Pinyol, Alejandra Mart{\'\i}nez-Trillos, et~al.
\newblock Exome sequencing identifies recurrent mutations of the splicing
  factor sf3b1 gene in chronic lymphocytic leukemia.
\newblock {\em Nature genetics}, 44(1):47--52, 2012.

\bibitem{Sael2012a}
Lee Sael and Daisuke Kihara.
\newblock {Constructing patch-based ligand-binding pocket database for
  predicting function of proteins.}
\newblock {\em BMC Bioinformatics}, 13(Suppl 2):S7, jan 2012.

\bibitem{shi2014rare}
Jianxin Shi, Xiaohong~R Yang, Bari Ballew, Melissa Rotunno, Donato Calista,
  Maria~Concetta Fargnoli, Paola Ghiorzo, Brigitte Bressac-de Paillerets,
  Eduardo Nagore, Marie~Francoise Avril, et~al.
\newblock Rare missense variants in pot1 predispose to familial cutaneous
  malignant melanoma.
\newblock {\em Nature genetics}, 46(5):482--486, 2014.

\bibitem{sunyaev1999psic}
Shamil~R Sunyaev, Frank Eisenhaber, Igor~V Rodchenkov, Birgit Eisenhaber,
  Vladimir~G Tumanyan, and Eugene~N Kuznetsov.
\newblock Psic: profile extraction from sequence alignments with
  position-specific counts of independent observations.
\newblock {\em Protein engineering}, 12(5):387--394, 1999.

\bibitem{tennessen2012evolution}
Jacob~A Tennessen, Abigail~W Bigham, Timothy~D O’Connor, Wenqing Fu, Eimear~E
  Kenny, Simon Gravel, Sean McGee, Ron Do, Xiaoming Liu, Goo Jun, et~al.
\newblock Evolution and functional impact of rare coding variation from deep
  sequencing of human exomes.
\newblock {\em science}, 337(6090):64--69, 2012.

\bibitem{touw2015series}
Wouter~G Touw, Coos Baakman, Jon Black, Tim~AH te~Beek, Elmar Krieger, Robbie~P
  Joosten, and Gert Vriend.
\newblock A series of pdb-related databanks for everyday needs.
\newblock {\em Nucleic acids research}, 43(D1):D364--D368, 2015.

\bibitem{witten2016data}
Ian~H Witten, Eibe Frank, Mark~A Hall, and Christopher~J Pal.
\newblock {\em Data Mining: Practical machine learning tools and techniques}.
\newblock Morgan Kaufmann, 2016.

\bibitem{Worth2011}
Catherine~L. Worth, Robert Preissner, and Tom~L. Blundell.
\newblock {SDM - A server for predicting effects of mutations on protein
  stability and malfunction}.
\newblock {\em Nucleic Acids Research}, 39(SUPPL. 2):215--222, 2011.

\bibitem{yan2011exome}
Xiao-Jing Yan, Jie Xu, Zhao-Hui Gu, Chun-Ming Pan, Gang Lu, Yang Shen, Jing-Yi
  Shi, Yong-Mei Zhu, Lin Tang, Xiao-Wei Zhang, et~al.
\newblock Exome sequencing identifies somatic mutations of dna
  methyltransferase gene dnmt3a in acute monocytic leukemia.
\newblock {\em Nature genetics}, 43(4):309, 2011.

\bibitem{Yang2002}
Haijuan Yang, Philip~D Jeffrey, Julie Miller, Elspeth Kinnucan, Yutong Sun,
  Nicolas~H Thoma, Ning Zheng, Phang-Lang Chen, Wen-Hwa Lee, and Nikola~P
  Pavletich.
\newblock {BRCA2 function in DNA binding and recombination from a
  BRCA2-DSS1-ssDNA structure.}
\newblock {\em Science (New York, N.Y.)}, 297(5588):1837--48, 2002.

\bibitem{yuan2006fastsnp}
Hsiang-Yu Yuan, Jen-Jie Chiou, Wen-Hsien Tseng, Chia-Hung Liu, Chuan-Kun Liu,
  Yi-Jung Lin, Hui-Hung Wang, Adam Yao, Yuan-Tsong Chen, and Chun-Nan Hsu.
\newblock Fastsnp: an always up-to-date and extendable service for snp function
  analysis and prioritization.
\newblock {\em Nucleic acids research}, 34(suppl 2):W635--W641, 2006.

\bibitem{Zhang2013a}
Feng Zhang, Cheng Hu, Yang Dong, Ming-Shen Lin, Jingyao Liu, Xinmei Jiang,
  Yubin Ge, and Yingjie Guo.
\newblock {The impact of V30A mutation on transthyretin protein structural
  stability and cytotoxicity against neuroblastoma cells.}
\newblock {\em Archives of biochemistry and biophysics}, 535(2):120--127, jul
  2013.

\end{thebibliography}

\end{document}